**Synthesis of RE123 high-$T_\text{c}$ superconductors with a high-entropy-alloy-type RE site**

Yuta Shukunami[1], Aichi Yamashita[1], Yosuke Goto[1], Yoshikazu Mizuguchi[1]*

1. *Department of Physics, Tokyo Metropolitan University, 1-1, Minami-osawa, Hachioji 192-0397, Japan.*

*Corresponding author: Yoshikazu Mizuguchi (mizugu@tmu.ac.jp)

Abstract

We have synthesized high-entropy-alloy-type (HEA-type) RE123 superconductors $Y_{0.28}Nd_{0.16}Sm_{0.18}Eu_{0.18}Gd_{0.20}Ba_2Cu_3O_{7-\delta}$ and $Y_{0.18}La_{0.24}Nd_{0.14}Sm_{0.14}Eu_{0.15}Gd_{0.15}Ba_2Cu_3O_{7-\delta}$ with a superconducting transition temperature ($T_\text{c}$) exceeding 90 K. From comparison between HEA-type and conventional RE123 samples, we found that the mixing entropy at the rare earth site does not affect $T_\text{c}$ and critical current density ($J_\text{c}$) of the RE123 samples. For all the examined samples, orthorhombicity was found to be an essential parameter for $T_\text{c}$ and $J_\text{c}$. In the regime near orthorhombic-tetragonal boundary, $J_\text{c}$ for the HEA-type sample was larger than that of conventional-type samples, which suggests that HEA-type RE123 may be useful to improve $J_\text{c}$ in the structure with low orthorhombicity.

1. Introduction

High-entropy alloys (HEAs) [1], which are defined as alloys containing five or more elements with a concentration between 5 to 35at%. Alloys synthesized with the criterion results in high configurational mixing entropy. HEAs have been typically studied in the field of structural materials; for example, stability in high temperature and/or extreme conditions are improved by increasing mixing entropy. Furthermore, in recent years, HEAs have been extensively studied in various fields of functional materials. One of the growing fields is HEA superconductors [2,3]. The HEA concept can be useful to develop new superconducting materials containing a HEA site and/or HEA-type layers. Figure 1 explains about the concept of possible design strategies of HEA-type superconductors. The schematic images of a crystal structure were prepared using VESTA software [4]. Figure 1A shows the structure of the simple HEA superconductor Ti-Zr-Hf-Nb-Ta [2], which is the firstly-reported HEA superconductor Ti-Zr-Hf-Nb-Ta with a transition temperature ($T_c$) of 7.3 K. A metal site, which is in a HEA state, is drawn with five different colors, indicating that five different elements randomly occupy the site with occupations indicated by the coloring ratio. Namely, the HEA superconductor is composed of a metal site, which is randomly occupied with five different elements of Ti, Zr, Hf, Nb, and Ta. Although the observed superconductivity in most HEA superconductors discovered so far has been characterized as a conventional phonon-mediated type, the field of HEA superconductors has been growing because of their notable features and possible material development. For example, high pressure measurements revealed that the superconductivity states in a HEA superconductor is robust at pressure under 190 GPa [5]. This fact suggests that HEA superconductors may be useful under extreme conditions. Superconductivity was recently observed in CsCl-type HEAs, Sc-Zr-Nb-Rh-Pd and Sc-Zr-Nb-Ta-Rh-Pd, as well [6]. Since a CsCl-type compound has two crystallographic sites. The observation of superconductivity in a CsCl-type system suggests that the concept of HEA can be expanded into superconducting compounds, which are not a simple alloy with single crystallographic site. However, so far, there are a few examples of *HEA-type compounds* which show superconductivity. Therefore, for further development of HEA-type compounds, understanding the influence of the presence of a HEA site in the crystal structure to the superconducting characteristics is quite important. Recently, we discovered a superconductor AgInSnPbBiTe$_5$ with a NaCl-type structure [7]. As shown in Fig. 1B, AgInSnPbBiTe$_5$ is composed of a metal (*M*) site (cationic site) occupied with $M$ = Ag, In, Sn, Pb, and Bi and a Te site (anionic site). Therefore, this material can be regarded as a compound containing both HEA site and non-HEA site. Since, the metal telluride has covalent *M*-Te bonds with ionic characteristics of $M^{2+}$ and $Te^{2-}$, HEA-like *M*-Te bond lengths are expected to be included in the telluride. Namely, highly disordered superconductor can be designed, and the disorder effects of the bonding states can be controlled by changing the mixing entropy.

Another concept of material design using a HEA site is introduction of a HEA site into layered materials. We have reported the synthesis and superconducting properties of $REO_{0.5}F_{0.5}BiS_2$ (RE: rare earth) with a HEA-type REO blocking layer [8]; the $REO_{0.5}F_{0.5}BiS_2$ is a typical $BiS_2$-based layered superconductor [9-14]. In the examined sample, the RE site is occupied with five different RE elements as shown in Fig. 1(c). Interestingly, the increase in mixed entropy at the RE site affects the local structure of the conducting layer ($BiS_2$ layer), and the superconducting properties have been improved [15]. The improvement of the superconducting properties by high mixing entropy was explained by the suppression of local disorder in the BiS superconducting planes [16-18]. Motivated by those mysterious results in the HEA-type $BiS_2$-based superconductors, we have decided to investigate the effect of the inclusion of HEA sites in the RE123 high-$T_c$ cuprates [19]. Here, we show that the RE site of the RE123 structure can be in HEA states by mixing more than five different RE elements. This study has been also motivated by the fact that the critical current density ($J_c$) of RE123 containing two or three RE is greater than those of pure RE123 samples [20]. A superconducting transition with a $T_c$ exceeding 90 K was observed for HEA-type samples as well. We show preliminary investigation of magnetic $J_c$ for prepared various RE123 polycrystalline samples.

2. Experimental details

We have synthesized polycrystalline samples of $REBa_2Cu_3O_{7-\delta}$. For the RE site, Y, La, Nd, Sm, Eu, and Gd were used. For all the samples, Y was contained in the composition, which is to reduce the number of testing composition. Pr was removed from the candidate elements because inclusion of Pr resulted in suppression of superconductivity in $REBa_2Cu_3O_{7-\delta}$ samples. In the starting nominal composition, RE elements were evenly mixed, except for sample #6 (See Table I), whose starting nominal ratio was Y : Sm = 0.2 : 0.8. Powders of barium carbonate, copper oxide, and RE oxides and or RE hydroxides were used for the synthesis. The polycrystalline samples were synthesized by a two-stage solid-phase reaction method, which has been established for typical $REBa_2Cu_3O_{7-\delta}$. The raw chemicals with a nominal compositional ratio of RE : Ba : Cu = 1 : 2 : 3 were well mixed and pelletized. In the first sintering, the pellet was heated at 930 °C for 20 h in air, followed by furnace cooling. The obtained sample was ground, mixed, and pelletized. The second sintering was performed at 930 °C for 8 h in air, and the furnace was cooled down to 350 °C in 12 h and kept for 18 h. Finally, the sample was furnace-cooled. We have tested various annealing temperatures of 300 °C, 350 °C, 400 °C, and 450° C, and found that annealing at 350 °C resulted in the best superconducting properties. The obtained samples were labeled #1–#15 as listed in Table I.

To characterize the purity and the crystal structure of the synthesized sample, powder X-ray diffraction (XRD) was performed on MiniFlex-600 (RIGAKU), which is equipped with a D/tex-Ultra detector, with a Cu-K$_\alpha$ radiation by a conventional $\theta$-$2\theta$ method. Rietveld refinement was performed using RIETAN-FP software [21] on the XRD patterns of the samples. The actual composition of the synthesized samples was investigated by energy dispersive X-ray spectroscopy (EDX) on a scanning electron microscope SEM [TM-3030 (Hitachi)]. The superconducting properties were investigated using a superconducting quantum interference device (SQUID) magnetometer on MPMS-3 (Quantum Design). Temperature dependence of magnetic susceptibility was measured after both zero-field cooling (ZFC) and field cooling (FC) with an applied field of about 10 Oe. The critical current density was estimated from the magnetic-field dependence of the magnetization (*M-H* curve) using the Bean London model [22].

3. Results and discussion

We have characterized the synthesized samples using EDX and XRD. The RE-site compositions obtained from EDX analyses are summarized in Table I. The analyzed compositions are close to the starting nominal compositions. For samples #12–#15, the RE site is occupied five or six RE elements. This condition satisfies the criterion of HEA; hence, we call those compounds *HEA-type RE123*.

As shown in Fig. 2, the powder XRD pattern for all the samples showed that a RE123 phase is a major phase. Although very small impurity peaks were observed, almost all the peaks were indexed using an orthorhombic space group of *Pmmm* (#47); the Miller indices are displayed in Fig. 2. To obtain crystal structure parameters, we performed Rietveld refinement on the XRD pattern for all the samples. Obtained in-plane lattice constants *a* and *b* and a reliability factor $R_{wp}$ are summarized in Table I. Typical result of the Rietveld refinement, which is for #14 (HEA-type RE123), is shown in Fig. 3. The refinement result shows that the peaks are nicely simulated by the orthorhombic model, and there is no clear phase separation and/or broadening of the XRD peak. On the basis of the results from EDX and XRD analyses, we have concluded that the RE site of the RE123 system can be homogeneously mixed with more than five RE elements, which makes the compound HEA-type RE123.

To investigate superconducting properties, we measured temperature dependence of magnetic susceptibility for all the samples. Except for #12 and #13, synthesized samples showed a superconducting transition. Typical temperature dependence of magnetic susceptibility for #14 is shown in Fig. 4(a). Diamagnetic signals corresponding to superconductivity are observed. $T_c$ of #14 (HEA-type RE123) is 93 K, which is comparable to conventional RE123 superconductors [19,22]. First, we plotted the estimated Tc for all the samples as a function of lattice constants *a*, *b*, and *c*, but no clear correlation between $T_c$ and lattice constants was observed. $T_c$ was also plotted as a function of number of RE elements, which is related to the mixing entropy at the RE site, we did not obtain clear correlation. Then, we plotted the estimated $T_c$ as a function of orthorhombicity parameter (*OP*), which is defined as $2|a-b| / (a+b)$, where *a* and *b* are lattice constants. As generally known, superconductivity of RE123 compounds is observed when the crystal structure is orthorhombic. With decreasing oxygen deficiency $\delta$, the orthorhombic crystal structure changes to tetragonal. $T_c$ of the orthorhombic phase decreases with approaching the orthorhombic-tetragonal boundary [23]. Therefore, the *OP* is a good scale for superconductivity in the RE123 system. As shown in Fig. 5(a), $T_c$ shows a clear correlation with *OP*. When $OP > 0.012$, $T_c$ of the synthesized samples is almost constant. However, $T_c$ decreases as *OP* decreases from 0.012, and superconductivity was not observed when $OP < 0.007$. The symbols in Fig. 5(a) indicate the number of RE element contained in the samples. We note that HEA-type RE123 samples also show superconductivity with a $T_c$ higher than 90 K. $T_c$ for the examined samples is depending on the orthorhombicity but not on mixing entropy.

For the samples in which superconductivity was observed, we measured magnetic field dependence of magnetization. The typical *M-H* curves for #14 at various temperatures are displayed in Fig. 4(b). From the *M-H* curves, we calculated magnetic $J_c$ using the Bean's model, which gives $J_c = 20\Delta M / B(1-B/3A)$ (A/cm$^2$), where *A* and *B* are determined by sample shape, and $\Delta M$ is obtained from the width of the *M-H* curve. The estimated magnetic $J_c$ for #14 is plotted in Fig. 4(c). A self-field $J_c$ for #14 exceeds 20 kA/cm$^2$, and $J_c$ at 1 T is 11.7 kA/cm$^2$. To compare $J_c$ of all the superconducting samples, $J_c$ at 1 T is summarized in Table I and plotted in Fig. 5(b) as a function of *OP*. As well as in Fig. 5(a), the symbols in Fig. 5(b) also indicate the number of RE element contained in the samples. $J_c$ also shows correlation with *OP*; with increasing *OP*, $J_c$ increases. Notably, the presence of HEA-type RE site in RE123 superconductors does not affect this trend on $J_c$-*OP*. Interestingly, when $OP = 0.012$–0.015, the largest $J_c$ is observed in #15 with six RE elements. Although we need further investigations on the relationship between composition, crystal structure including local disorder, mixing entropy, and superconducting properties, our present results suggest that designing HEA-type RE123 superconductors will be a useful strategy to increase $J_c$ for materials with lower orthorhombicity.

If we could synthesize tetragonal (or low-orthorhombicity) RE123 superconductor samples with a $T_c$ higher than 90 K and a $J_c$ larger than the current largest record of $J_c$ in RE123 systems, it will be useful for practical applications like superconducting wires.

4. Conclusion

We have synthesized HEA-type RE123 superconductor samples of $Y_{0.28}Nd_{0.16}Sm_{0.18}Eu_{0.18}Gd_{0.20}Ba_2Cu_3O_{7-\delta}$ (#14) and $Y_{0.18}La_{0.24}Nd_{0.14}Sm_{0.14}Eu_{0.15}Gd_{0.15}Ba_2Cu_3O_{7-\delta}$ (#15) with a $T_c$ exceeding 90 K. For comparison, various RE123 compounds with different number of containing RE elements, from one to six, have been investigated. We found that the mixing entropy at the RE site does not affect $T_c$ and $J_c$ of the RE123 samples. Commonly, orthorhombicity is an essential parameter for $T_c$ and $J_c$ in the RE123 samples. From estimation of magnetic $J_c$ at 1 T, we suggested that HEA-type RE123 superconductor may have a larger $J_c$ than conventional compositions, which can be a possible strategy to design large-$J_c$ RE123 with tetragonal (or low-orthorhombicity) structure, which is suitable for practical applications like superconducting wires.

**Acknowledgements**

The authors thank O. Miura, K. Hoshi, R. Sogabe, T. Ozaki, A. Miura, and M. Nagao for their experimental supports and fruitful discussion. This work was partly supported by Grants-in-aid for scientific research (Nos.: 15H05886 and 18KK0076) and the Advanced Research Program under the Human Resources Funds of Tokyo (Grant Number: H31-1).

Table I. Composition for the RE site, in-plane lattice constants ($a$ and $b$), reliability factor $R_{wp}$ of Rietveld refinement, $T_c$, and magnetic $J_c$ at 1 T for RE123 samples with different RE-site occupancy.

| Label | Ratio for RE site (EDX) | $a$ (Å) | $b$ (Å) | $R_{wp}$ (%) | $T_c$ (K) | $J_c$ (1T) (kA/cm$^2$) |
|---|---|---|---|---|---|---|
| #1 | Y | 3.81756(10) | 3.8843(2) | 6.1 | 93 | 28.3 |
| #2 | $Y_{0.50}La_{0.50}$ | 3.8596(2) | 3.9137(3) | 8.8 | 86 | 3.0 |
| #3 | $Y_{0.57}Nd_{0.43}$ | 3.84687(14) | 3.9088(2) | 10.4 | 92 | 20.1 |
| #4 | $Y_{0.58}Sm_{0.42}$ | 3.83446(11) | 3.90067(10) | 8.2 | 93 | 31.2 |
| #5 | $Sm_{0.48}Eu_{0.42}$ | 3.8690(2) | 3.9055(2) | 6.9 | 40 | 1.1 |
| #6 | $Y_{0.19}Sm_{0.81}$ | 3.85384(14) | 3.9090(2) | 6.7 | 45 | 3.5 |
| #7 | $Y_{0.41}La_{0.32}Nd_{0.27}$ | 3.8679(2) | 3.9192(3) | 9.1 | 91 | 5.2 |
| #8 | $Y_{0.38}La_{0.35}Sm_{0.27}$ | 3.8583(2) | 3.9104(3) | 7.8 | 92 | 4.8 |
| #9 | $Y_{0.42}Nd_{0.30}Sm_{0.28}$ | 3.8542(2) | 3.91413(14) | 11.8 | 93 | 14.4 |
| #10 | $Y_{0.39}Sm_{0.30}Eu_{0.31}$ | 3.84034(10) | 3.9024(2) | 5.8 | 93 | 32.9 |
| #11 | $Y_{0.26}La_{0.28}Nd_{0.22}Sm_{0.24}$ | 3.8658(2) | 3.9187(3) | 9.1 | 92 | 5.1 |
| #12 | $Y_{0.23}La_{0.25}Nd_{0.15}Sm_{0.19}Eu_{0.18}$ | 3.8821(2) | 3.8976(2) | 5.8 | non-SC | non-SC |
| #13 | $Y_{0.23}La_{0.23}Nd_{0.17}Sm_{0.18}Gd_{0.19}$ | 3.8763(2) | 3.9024(2) | 6.1 | non-SC | non-SC |
| #14 | $Y_{0.28}Nd_{0.16}Sm_{0.18}Eu_{0.18}Gd_{0.20}$ | 3.84538(9) | 3.90434(9) | 4.9 | 93 | 11.7 |
| #15 | $Y_{0.18}La_{0.24}Nd_{0.14}Sm_{0.14}Eu_{0.15}Gd_{0.15}$ | 3.86362(13) | 3.91251(13) | 8.7 | 93 | 8.1 |

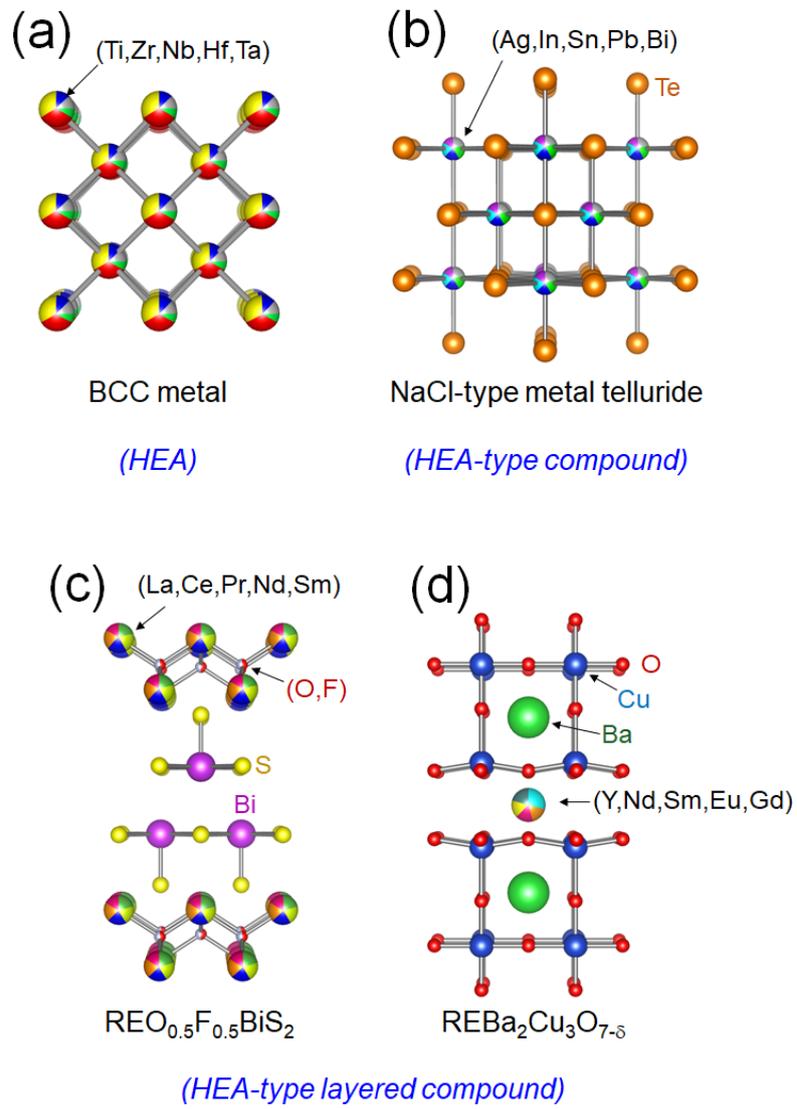

Figure 1. Crystal structure images of various types of HEA-type superconductors.

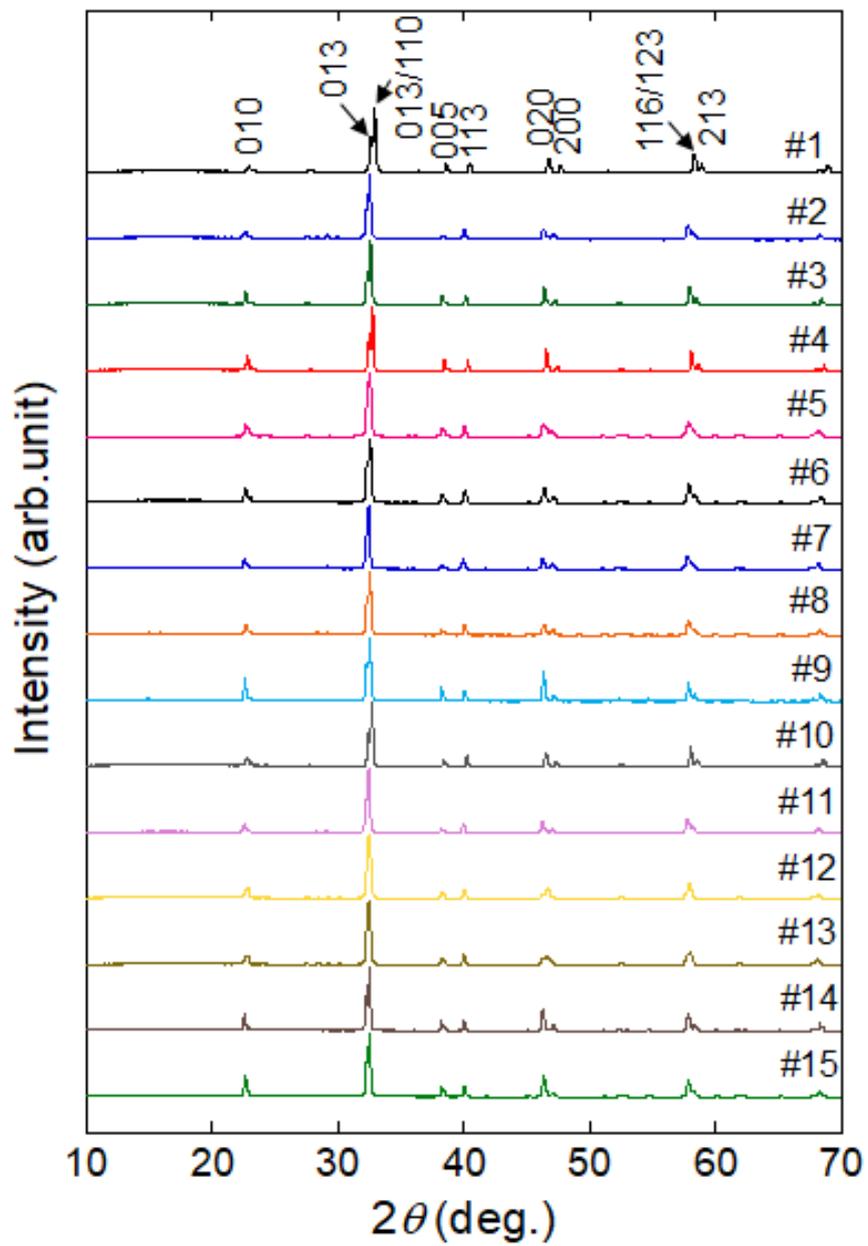

Figure 2. Powder XRD patterns for all the samples. The numbers in the figure are Miller indices.

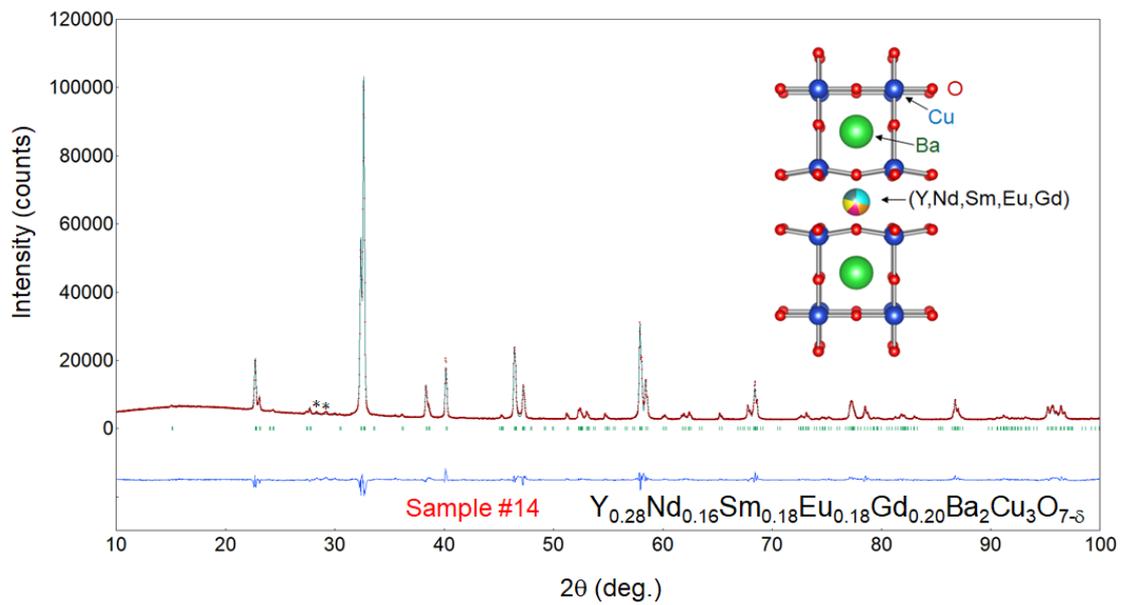

Figure 3. Powder XRD pattern and Rietveld refinement result for sample #14. A residual curve is shown in the bottom. Impurity peaks are indicated with asterisks.

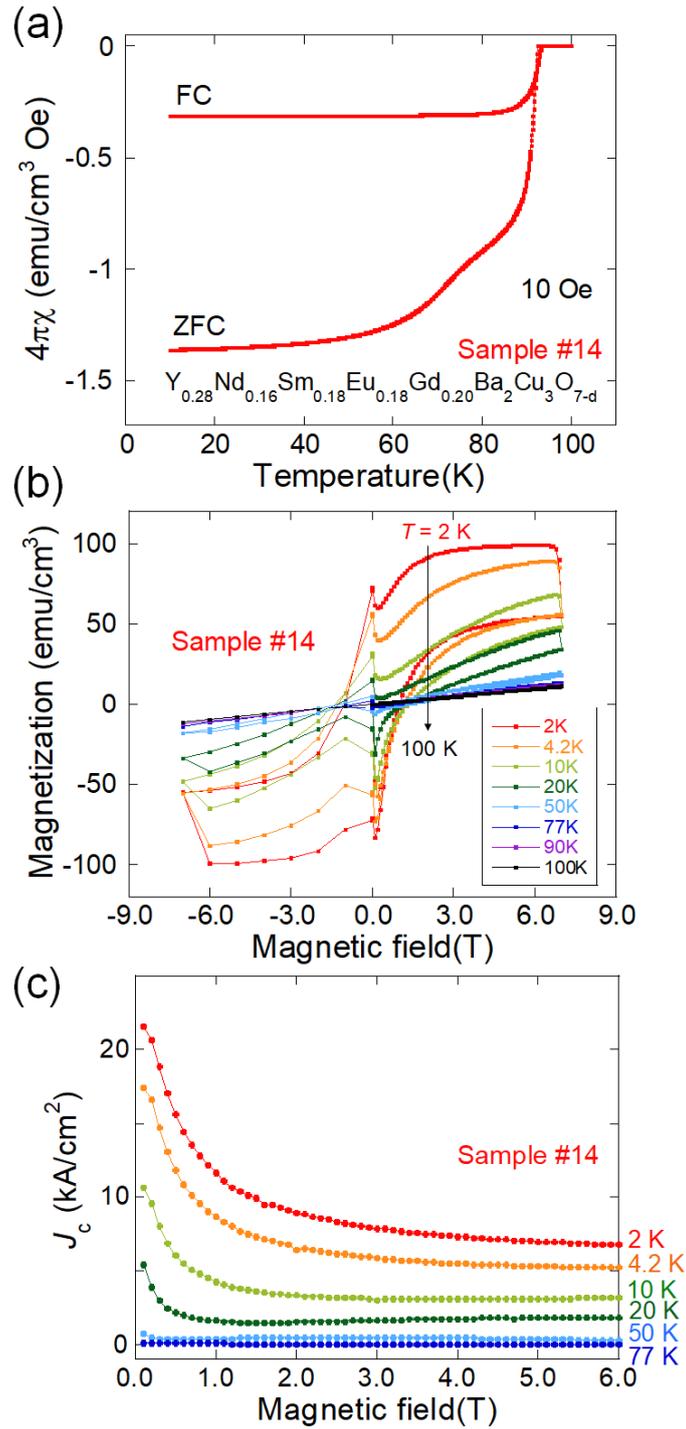

Figure 4. (a) Temperature dependence of magnetic susceptibility ($4\pi\chi$) for sample #14. (b) Magnetic field dependences of magnetization at $T$ = 2–100 K. (c) Magnetic field dependences of magnetic $J_c$ at $T$ = 2–77 K estimated using Bean's model [22].

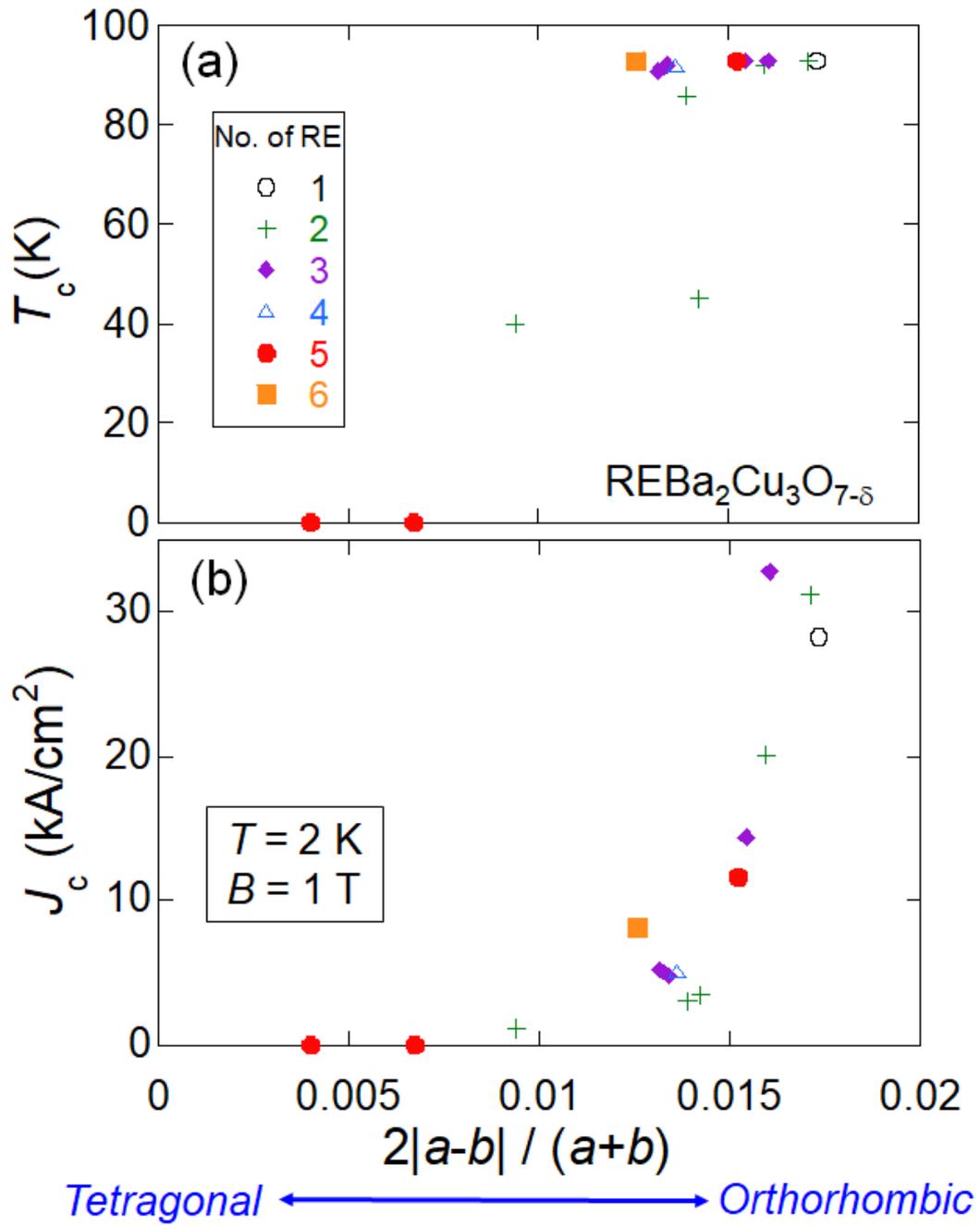

Figure 5. $T_c$ and $J_c$ are plotted as a function of an orthorhombicity parameter $OP = [2|a-b|/(a+b)]$, which is calculated from in-plane lattice constants obtained by Rietveld analyses.